\documentclass[amsmath, amssymb,10pt, aps, prl, twocolumn, notitlepage, longbibliography, superscriptaddress]{revtex4-2}
\usepackage{graphicx}
\usepackage{amsmath}
\usepackage{hyperref}
\usepackage{xcolor}

\begin{document}

\title{A footprint of zero-point entropy in higher-temperature magnetic thermodynamics}

\author{Sergey Syzranov}
\author{Arthur P. Ramirez}
\affiliation{Physics Department, University of California Santa Cruz, Santa Cruz, California 95064, USA}

\begin{abstract}
Identifying extensively degenerate zero-temperature states is key in characterizing spin-liquid-candidate materials and spin ices. 
In experiments, finding zero-point entropy (ZPE) is often attempted by measuring
the entropy released by a material when cooled down from very high to very low
temperatures. Such investigations are often unreliable and lead to controversial results because accessible temperatures may be insufficient to accurately capture essential 
low- and high-temperature features of magnetic materials.
The purpose of this paper is to point out a simple, easily accessible
signature of nonzero ZPE: the Maxwell's relation 
$\left(\partial S/\partial H\right)_T = \left(\partial M/\partial T\right)_H$
can appear violated if a vanishing ZPE is assumed incorrectly. This relation can further be used for estimating the ZPE.
In many materials below characteristic temperatures, the criterion
of non-vanishing ZPE has a particularly simple form: $\left(\frac{\partial C}{\partial H}\right)_T\left(\frac{\partial M}{\partial T}\right)_H<0$.
We discuss these effects and the ZPE signature in the benchmark test case of the 
well-studied spin ice $Dy_2Ti_2O_7$.
\end{abstract}

\maketitle

Determining whether a material possesses an extensively degenerate ground state
(zero-point-entropy) (ZPE) is key to identifying and characterizing spin ices and candidate spin-liquid materials. 
A common experimental strategy for estimating zero-point entropy (ZPE) is to measure the entropy
\begin{align}
    S_\text{E}=\int_0^\infty \frac{C(T)}{T} dT
    \label{ExcitationalEntropy}
\end{align}
released when the material is cooled down from very
high temperatures to very low temperatures and then compare it with 
the value $S_\text{expected}=N \ln (2s+1)$ that a spin system is expected to 
release in the absence of the ground-state degeneracy.
Insufficient released entropy, $S_\text{E}<S_\text{expected}$ signals
the residual entropy associated with an extensively degenerate ground state.

This strategy has been used by one of us~\cite{Ramirez1999} to observe the residual entropy in the spin-ice~\cite{Harris1997} compound $Dy_2Ti_2O_7$.
The measured ground-state entropy $S_\text{res}=S_\text{expected}-S_\text{E}$ matched closely the theoretical expectation~\cite{KnolleMoessner:review,Pauling1945}.

Based on the same approach, the existence of quantum spin liquids (QSLs) has been suggested
in a number of geometrically frustrated (GF) magnetic materials (see, for example, Refs.~\cite{Fenner:ZPE,KumarMahajan:triangular,NirmalaMatsuda:ZPE,Dai:CeZrO,Gaulin:CeZrO,Lee:BSZCGO}). This approach, however, is unreliable and often gives controversial results, owing both to the limited accessible temperature range and to the experimental challenge of measuring $C(T)$ accurately.

\begin{figure}
    \centering
    \includegraphics[width=0.9\linewidth]{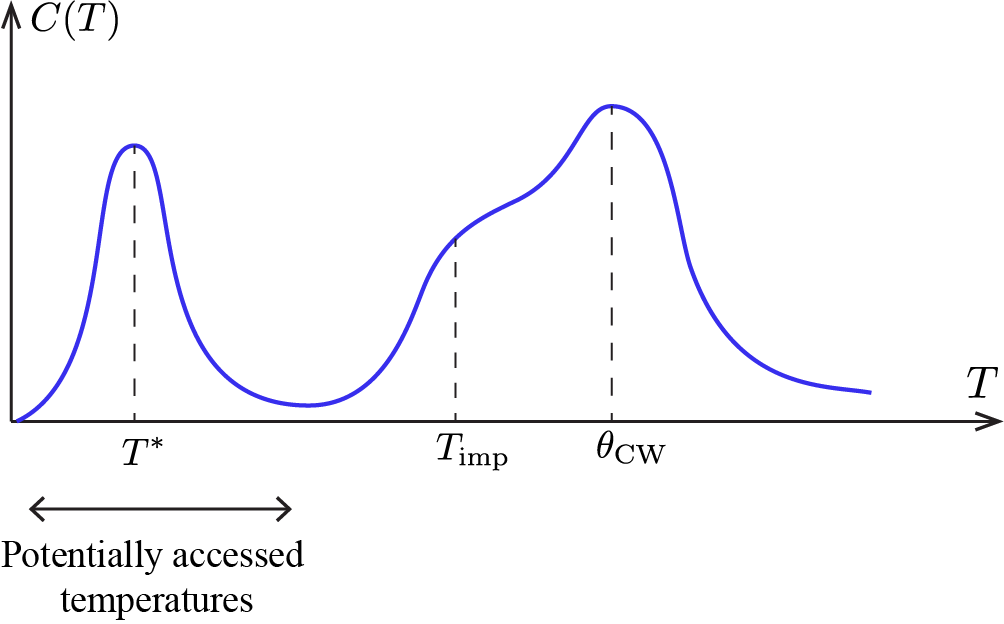}
    \caption{ \label{fig:Cplot}
    Common behavior of heat capacity in geometrically frustrated magnetic materials. The heat capacity $C(T)$ in such materials exhibits, in general, a peak at a temperature $T$ of the order of the Curie-Weiss
    temperature $\theta_\text{CW}$ and another peak at lower temperature 
    $T^*$ (the ``hidden energy scale''~\cite{Syzranov:HiddenEnergy,RamirezSyzranov:review,Popp:TwoPeak}).
    Randomly located magnetic and non-magnetic impurities can lead to the emergence of additional 
    peaks in $C(T)$ at temperatures $T_\text{imp}$ set by the impurity concentrations~\cite{Sedik:anomaly,Tari:book}.}
\end{figure}

It has been shown, in particular, that GF magnets exhibit two peaks in the behavior
of heat capacity $C(T)$ that may be well separated in temperature (see Fig.~\ref{fig:Cplot})~\cite{Elser:KHAF,ElstnerYoung:kagome,IsodaNakano:XXZ,SchnackSchulenberg:KHAF,Ulaga:EasyAxis,Popp:TwoPeak,RamirezSyzranov:review}. Experimentally accessed temperature ranges are usually insufficient to capture both peaks~\cite{Popp:TwoPeak,Zhu:entropies}, often missing either the lower-temperature or the higher-temperature peak, with the missing entropy 
beyond the accessed temperature range sometimes
mistakenly attributed to the ZPE.

{\it A Footprint of ZPE in higher-temperature thermodynamics.}
The goal of this letter is to point out a simple and readily measurable signature of nonzero ZPE in magnetic materials.
Specifically, the Maxwell relation (in Gaussian CGS units)
\begin{align}
\left(\frac{\partial S}{\partial H}\right)_T = \left(\frac{\partial M}{\partial T}\right)_H
\label{MR}
\end{align}
can seem to fail when one incorrectly assumes the vanishing of the  ZPE. Verifying ZPE
can be further simplified in certain materials: if the derivatives 
$\left(\frac{\partial C}{\partial H}\right)_T$ and $\left(\frac{\partial M}{\partial T}\right)_H$ have different signs below the lowest characteristic energy scale, 
such a material has a non-vanishing ZPE.
We examine these effects in the benchmark spin-ice system $Dy_2Ti_2O_7$.

{\it Excitational and ground-state entropies.}
The total entropy of a magnetic material can be represented as a sum of the ground-state entropy $S_0$ and 
the ``excitational'' entropy $S_E$ given by Eq.~\eqref{ExcitationalEntropy}:
\begin{equation}
S(T,H)=S_0(H)+S_E(T,H).
\end{equation}
Measurements of heat capacity $C(T)$ in a limited temperature interval allows one to only track changes of the excitational entropy $S_E(T,H)$. If zero ground-state entropy is assumed, $S_0=0$,
the Maxwell relation~\eqref{MR} will appear violated in a material with an extensively degenerate ground state, due to the change of the ground-state entropy with the magnetic field.
As we discuss below, this violation is especially vivid when the two sides of Eq.~\eqref{MR}
have opposite signs.

Because an external magnetic field $H$
typically promotes spin ordering in the ground state, it decreases $S_0(H)$.
Utilizing Eq.~\eqref{MR}, this leads to the criterion 
\begin{equation}
\left(\frac{\partial M}{\partial T}\right)_H<\int_0^T \frac{\partial}{\partial H}\frac{C(T,H)}{T}\,dT
\label{CriterionIntermediateSimplicity}
\end{equation}
of the non-vanishing ZPE. Condition~\eqref{CriterionIntermediateSimplicity} 
does not require measuring the heat capacity $C(T)$ in a temperature interval from the very lowest
to the highest energy scales in the material and is thus easier to verify than
obtaining the residual entropy $S_\text{res}=S_\text{expected}-S_\text{E}$ as 
the difference between the expected and excitational entropies.

Such a verification, however, may still require accessing rather low-temperature features
of the heat capacity. Detecting ZPE can be further simplified in many magnetic materials
if the experiment has access to temperatures just below the lowest energy scale in the material, such as the ``hidden energy scale''~\cite{Syzranov:HiddenEnergy,RamirezSyzranov:review,Popp:TwoPeak}
$T^*$ (cf. Fig.~\ref{fig:Cplot}). 
It can be assumed that thermodynamic observables do not change their behavior qualitatively, and,
for example, the quantity $\frac{\partial}{\partial H}\frac{C(T,H)}{T}$ does not change sign below the lowest characteristic energy scale.

Under this rather generic assumption, the existence of nonzero ZPE is guaranteed by the condition
\begin{align}
\left(\frac{\partial C}{\partial H}\right)_T\left(\frac{\partial M}{\partial T}\right)_H<0.
\label{SimplestCriterion}
\end{align}
This condition can be viewed as a violation of the Maxwell relation~\eqref{MR},
if zero ground-state entropy is assumed, due to the 
two parts of the equation having different signs.
Because it is sufficient to measure the magnetic susceptibility and heat capacity at only one value of the magnetic field $H$ and temperature $T$ to verify
the condition Eq.~\eqref{SimplestCriterion}, it constitutes a exceptionally simple and reliable 
criterion of ZPE.

{\it Test case: spin ice $Dy_2Ti_2O_7$.} The spin-ice material $Dy_2Ti_2O_7$
is a perfect benchmark system for demonstrating higher-temperature ZPE signatures. 
The residual entropy $S_\text{res}=S_\text{E}-S_\text{expected}$ in this material
matches~\cite{Ramirez1999} the theoretical value~\cite{KnolleMoessner:review}, which is closely approximated by the Pauling entropy~\cite{Pauling1945} $S_\text{Pauling}=\ln(3/2)/2$ (per spin, in units of $R$).

The heat capacity $C(T)$ of $Dy_2Ti_2O_7$ has been shown~\cite{Ramirez1999} to exhibit
a peak at a temperature of $T\approx 0.7K$ and, at temperatures below this peak,
decreases when the field is increased
from $H=0T$ to $H=0.5T$, thus demonstrating $\left(\frac{\partial C}{\partial H}\right)_T<0$
at a certain field in the respective interval.

Measurements of the magnetic susceptibility, $\chi(T)$, of Dy$_2$Ti$_2$O$_7$ show a monotonic decrease with lowering temperature below a peak in $\chi(T)$, a value closely matching the peak temperature of $C(T)$ \cite{Matsuhira2001,Yaraskavitch2012,Takatsu2013}, which leads to the positivity of the derivative
$\left(\frac{\partial M}{\partial T}\right)_H>0$ in the respective range of parameters.

Different signs of the derivatives $\left(\frac{\partial C}{\partial H}\right)_T<0$
and $\left(\frac{\partial M}{\partial T}\right)_H>0$, observed below the lowest characteristic temperature, which in the case of $Dy_2Ti_2O_7$ 
is the $C(T)$ peak temperature, guarantees the existence of a non-vanishing
ZPE in this material.

{\it Acknowledgements.}  This work has been supported by the NSF grant DMR2218130. We are grateful to P.~Popp, M.~Sedik, and S.~Zhu for collaboration  on related topics.


%

\end{document}